\documentclass[manuscript]{aastex6}

\usepackage{breakurl}
\shorttitle{SUNWARD ALFV\'ENIC  FLUCTUATIONS}
\shortauthors{LI ET AL.}

\begin{document}


\title{Sunward-propagating Alfv\'enic fluctuations observed in the heliosphere}



\author{Hui Li \altaffilmark{1}, Chi Wang \altaffilmark{1}, John W. Belcher \altaffilmark{2}, Jiansen He \altaffilmark{3}, and John D. Richardson \altaffilmark{2}}


\altaffiltext{1}{State Key Laboratory of Space Weather, National Space Science Center, CAS, Beijing, 100190, China; \url{hli@spaceweather.ac.cn}}

\altaffiltext{2}{Kavli Institute for Astrophysics and Space Research, Massachusetts Institute of Technology, Cambridge, MA, USA}

\altaffiltext{3}{School of Earth and Space Sciences, Peking University, Beijing, 100871, China}

\begin{abstract}
The mixture/interaction of anti-sunward propagating Alfv\'enic fluctuations (AFs) and sunward-propagating Alfv\'enic fluctuations (SAFs) is believed to result in the decrease of Alfv\'enicity of solar wind fluctuations with increasing heliocentric distance. However, SAFs are rarely observed at 1 AU and solar wind AFs are found to be generally outward. Using the measurements from \textit{Voyager} 2 and \textit{Wind},  we perform a statistical survey of SAFs in the heliosphere inside 6 AU. We first report two SAF events observed by \textit{Voyager} 2. One is in the anti-sunward magnetic sector with a strong positive correlation between the fluctuations of magnetic field and solar wind velocity. The other one is in the sunward magnetic sector with a strong negative magnetic field--velocity correlation. Statistically, the percentage of SAFs increases gradually with heliocentric distance, from about 2.7\% at 1.0 AU to about 8.7\% at 5.5 AU. These results provide new clues to help understand the generation mechanism of SAFs.

\end{abstract}

\keywords{magnetic fields -- magnetohydrodynamics (MHD) -- plasmas -- solar wind -- turbulence -- waves}



\section{Introduction} 
\label{sec:intro}

Since the 1960s, the frequent presence of Alfv\'en waves or Alfv\'enic fluctuations has been identified from in situ observations of solar wind fluctuations over the radial range from 0.3 to 20 AU and from the ecliptic plane to high-latitudes \citep[see][and references therein]{Belcher and Davis 1971,Burlaga 1971,Volk 1975,Tu and Marsch 1995,Yang and Chao 2013}. The Alfv\'enic fluctuations (AFs) mostly originate from the Sun and thus mostly propagate in the anti-sunward direction. In general, the flow velocity fluctuations are negatively/positively correlated with magnetic field fluctuations in the anti-sunward (sunward) heliospheric magnetic field sector.

The interactions of counter-propagating AFs are thought to be an important source of solar wind plasma heating and decreasing Alfv\'enicity \citep{Burlaga and Turner 1976,van der Holst et al 2014}. However, only a few clear events of sunward-propagating Alfv\'enic fluctuations (SAFs) are reported in the literature. \citet{Roberts et al 1987} and \citet{Marsch 1991} found that discrete SAFs are rare in the pristine solar wind at 1 AU. \citet{Gosling et al 2009,Gosling et al 2011} performed a limited search for the signatures of discrete SAFs in the \textit{ACE} and \textit{Wind} data and identified a limited number of periods with SAFs. They found that  SAFs were found only (1) in events associated with back-streaming ions from the Earth's bow shock, (2) immediately up and down stream from reverse shocks associated with corotating interaction regions or interplanetary coronal mass ejections, and (3) in events identified as reconnection exhausts. Recently, \citet{Wang et al 2015} utilized a new criterion to identify the upstream-propagating Alfv\'enic intervals in the upstream region of the Earth's bow shock and found both upstream-propagating AFs with a power spectral bump due to the linear ion beam instability and upstream-propagating AFs with power law spectra due to a nonlinear wave-wave interaction. \citet{He et al 2015a} later reported the first observation of SAFs in the solar wind at 1 AU in the region magnetically disconnected from the Earth's bow shock. 

\citet{Bruno et al 1997} used Els\"{a}sser variables to represent the anti-sunward ($\delta\textbf{Z}^+$) and sunward ($\delta\textbf{Z}^-$) sense of propagation with respect to the Sun, and discussed the nature of the sunward component of AFs at 0.3 AU. \citet{Bavassano et al 2000,Bavassano et al 2001} later adopted the similar analysis methodology and studied the evolution of the anti-sunward and sunward components of AFs in the solar wind both at high-latitudes and in the ecliptic plane. However, these authors acknowledge doubt as to whether or not $\delta\textbf{Z}^-$ fluctuations, at scales smaller than 1 hour, represent SAFs. For example, $\delta\textbf{Z}^-$ can be the sunward-propagating quasi-perpendicular slow-mode waves, which have been clearly identified by \citet{He et al 2015b} in the compressible solar wind turbulence. The power spectral density of $\delta\textbf{Z}^-$ in 2D wave-vector space, which is derived using a spectral tomography method as introduced by \citet{He et al 2013}, shows a quite different distribution from that of $\delta\textbf{Z}^+$, with the former being more quasi-perpendicular and dominated by magnetic field fluctuations \citep{Yan et al 2016}. Such differences suggest that $\delta\textbf{Z}^-$ may not necessarily be the SAFs as previously conceived.

To our knowledge,  no case of SAFs beyond 1 AU has previously been reported. The statistical properties of SAFs with different heliocentric distances are still unknown. Using measurements from the \textit{Voyager} 2 and \textit{Wind} spacecraft, we present the first two clear cases of SAFs observed beyond 1 AU, far from the foreshock regions of planets. Based on a statistical survey, we find that the percentage of SAFs increases gradually with heliocentric distance.

\section{Data Set and Methodology}
\label{sec:data}

The present analysis uses the solar wind plasma and magnetic field data from \textit{Voyager} 2 during 1977 and 1979 and from \textit{Wind} during 1998 and 1999. These two time periods are both in the rising phase of the solar cycle.

\textit{Voyager} 2 was launched on August 20, 1977 and continues to explore the heliosphere. Plasma data from the PLS instrument \citep{Bridge et al 1977} have a sampling period of 12 seconds inside of 3 AU, of 96 seconds from 3 to 6 AU, and of 192 seconds beyond 6 AU. The magnetic field data are from the MAG instrument \citep{Behannon et al 1977};  we use averages of the field data over the plasma measurement cycle. During 1977 and 1979, \textit{Voyager} 2 was in the ecliptic plane inside of 6 AU. Thus, the resolution of  \textit{Voyager} 2 data combined plasma and magnetic field are chosen to be 48 or 96 seconds, which is adequate for studying AFs with period larger than 100 or 200 seconds. The data during the Jupiter observation phase have been excluded to eliminate the interference of the planetary bow shock, since SAFs may be produced by back-streaming ions from the planetary bow shock \citep[e.g.,][]{Gosling et al 2009,Wang et al 2015}.

The \textit{Wind} spacecraft was launched on 1 November 1994. The 3-D Plasma and Energetic Particle (3DP) instrument  on \textit{Wind} provides full three-dimensional measurements with high sensitivity of solar wind plasma \citep{Lin et al 1995}. The Magnetic Field Investigation on \textit{Wind} consists of a dual triaxial fluxgate DC magnetometer \citep{Lepping et al 1995}. The time resolution used here is 3 seconds. To compare with \textit{Voyager} 2 results in a statistical sense, \textit{Wind} data from 1998 to 1999 are used, which corresponds to the same phase of the solar cycle as the \textit{Voyager} 2 data used in this study.

The approach of \citet{Li et al 2016} is used to identify interplanetary AFs. Compared to conventional Wal\'en test methods, the deHoffmann-Teller (HT) frame and the background magnetic field are not needed to be determined in advance. Thus, the uncertainties introduced in the determinations of these two parameters could be reduced. We here use the band-pass filtered signals of the plasma velocity and magnetic field observations, instead of the original data sets, to check the Wal\'en relation. The property of pure AFs in the frequency domain can be accordingly obtained for each band-passed signal as follows:
\begin{linenomath}
\begin{equation}
\label{wr4}
   \delta \mathbf{V}_i = \pm \delta \mathbf{V}_\mathrm{Ai}
\end{equation}
\end{linenomath}
Here, $\delta \mathbf{V}_i$ and $\delta \mathbf{V}_\mathrm{Ai}$ represent the band-passed $\mathbf{V}$ (solar wind velocity) and $\mathbf{V}_\mathrm{A}$ (local Alfv\'en velocity) with the $ith$ filter, respectively. The sign $-/+$ denotes respectively propagation parallel and anti-parallel to the background magnetic field.

In the literature, several parameters are defined to represent the Alfv\'enicity, such as the Alfv\'en ratio, the Wal\'en slope, the normalized cross helicity, the normalized residual energy, and the velocity-magnetic field correlation coefficient \citep[see][and references therein]{Li et al 2016}. However, each parameter has its own limitations. For example, the Alfv\'en ratio, the normalized cross helicity and the normalized residual energy themselves do not necessarily require that the fluctuations of velocity and magnetic field correlate well. A good velocity--magnetic field correlation coefficient does not guarantee that the fluctuations match the Wal\'en relation. Thus, we use a more reliable quantity proposed by \citep{Li et al 2016}, $E_{rr}$, to assess the goodness of the Wal\'en test and the degree of Alfv\'enicity.

For each time series, we calculate $\delta \mathbf{V}_i$ and $\delta \mathbf{V}_\mathrm{Ai}$ for different frequency filters. For each filtered data set, we calculate the following eight parameters: 1) $\left|\left|\gamma_c\right|-1\right|$; 2) $\left|\left|\gamma_{cx}\right|-1\right|$; 3) $\left|\left|\gamma_{cy}\right|-1\right|$; 4) $\left|\left|\gamma_{cz}\right|-1\right|$; 5) $\left|\frac{\sigma_{\delta \mathbf{V}}}{\sigma_{\delta\mathbf{V}_\mathrm{A}}}-1\right|$; 6) $\left|\frac{\sigma_{\delta V_x}}{\sigma_{\delta V_{\mathrm{A}x}}}-1\right|$; 7) $\left|\frac{\sigma_{\delta V_y}}{\sigma_{\delta V_{\mathrm{A}y}}}-1\right|$; 8) $\left|\frac{\sigma_{\delta V_z}}{\sigma_{\delta V_{\mathrm{A}z}}}-1\right|$. Here, $\gamma_c$ is the correlation coefficient between all the components of $\delta \mathbf{V}$ and $\delta \mathbf{V}_\mathrm{A}$, $\sigma_{\delta \mathbf{V}}$ represents the standard deviation of all the components of $\delta \mathbf{V}$, and $\sigma_{\delta \mathbf{V}_\mathrm{A}}$ represents the standard deviation of all the components of $\delta \mathbf{V}_\mathrm{A}$. The terms with subscript $x, y,$ and $z$ are for the $x, y$, and $z$ components. The parameter $E_{rr}$ is the average value for these eight parameters.

In this study, we use a moving window with a width of 1 hour and a moving step of 5 min to calculate $E_{rr}$. The fluctuations in intervals with $E_{rr} <$ 0.15 are regarded as AFs. For 3-s \textit{Wind} data, the filters are chosen to be 10 -- 15 s, 15 -- 25 s, 25 -- 40 s, 40 -- 60 s, 60 -- 100 s, 100 -- 160 s, 160 -- 250 s, 250 -- 400 s, 400 -- 630 s, and 630 -- 1000 s. For 48-s \textit{Voyager} 2 data, the filters are chosen to be 100 -- 135 s, 135 -- 180 s, 180 -- 250 s, 250 -- 330 s, 330 -- 450 s, 450 -- 600 s, 600 -- 810 s, 810 -- 1100 s, 1100 -- 1480 s, and 1480 -- 2000 s. For 96-s \textit{Voyager 2} data, the filters are chosen to be 200 -- 250 s, 250 -- 320 s, 320 -- 400 s, 400 -- 500 s, 500 -- 630 s, 630 -- 800 s, 800 -- 1000 s, 1000 -- 1260 s, 1260 -- 1580 s, and 1580 -- 2000 s.

The wave propagation direction is determined according to the direction of the background magnetic field. However, as an unmeasurable parameter, the background magnetic field is hard to be determined accurately. The mean magnetic field is often assumed to be a proxy. But it is difficult to select time intervals over which the averages should be taken. Here we assume that the mean value of the low-passed magnetic field ($>$ 2000 s) equals the background magnetic field. The time period of interplanetary AFs varies from several minutes to a few hours. The largest time period of the filter we are concerned with is less than 2000 s. This assumption smooths out most wave effects. Note that, the background magnetic field is only used to justify whether an AF is sunward or anti-sunward. The uncertainties in determining the background magnetic field do not significantly affect our statistical conclusions. For AFs, the positively (negatively) correlated fluctuations of flow velocity and magnetic field represent anti-parallel (parallel) propagation relative to the background magnetic field \citep{Alfven 1942,Burlaga 1971}. The criteria of $\left|\left<B_R\right>\right|/\left<\left|B\right|\right> > 0.5$, $\left<B_R\right>\cdot CC > 0$ (for \textit{Voyager} 2 data in the RTN coordinate system) and $\left|\left<B_X\right>\right|/\left<\left|B\right|\right> > 0.5$, $\left<B_X\right>\cdot CC < 0$ (for \textit{Wind} data in the GSE coordinate system) are used to select the intervals of SAFs in the solar wind frame, where the brackets represent the mean value and $CC$ is the velocity - magnetic field correlation coefficient. Such intervals are defined as potential SAFs, which are a subset of AFs in our work.

Local bending of the interplanetary magnetic field line can make determination of the sunward direction difficult \citep[see][]{He et al 2011}. \citet{Wang et al 2015} defined the intervals with waves propagating in a direction opposite to that of the observed strahl electron outflow to be sunward propagating. Unfortunately, the strahl electron information is not available for \textit{Voyager} 2 data, so an additional criterion is adopted to reduce the interference of magnetic field bending. The angles between our determinations of the background magnetic field of the potential SAFs and the upstream/downstream solar wind are calculated. If these two angles are both less than 60$^\circ$,  such potential SAFs are defined as SAFs. Otherwise, the potential SAFs are defined as pseudo-SAFs caused by the local bending of magnetic field lines, and are excluded from our study.

Based on the criteria described above, we searched the AFs and SAFs from \textit{Voyager} 2 data during 1977 and 1979 and from \textit{Wind} data during 1998 and 1999. These two time intervals are in similar phases of the solar cycle, both before solar maximum. 

\section{Results}
\label{sec:result}
\subsection{Two SAF Events}

\begin{figure}
\centering
\noindent\includegraphics[width=30pc]{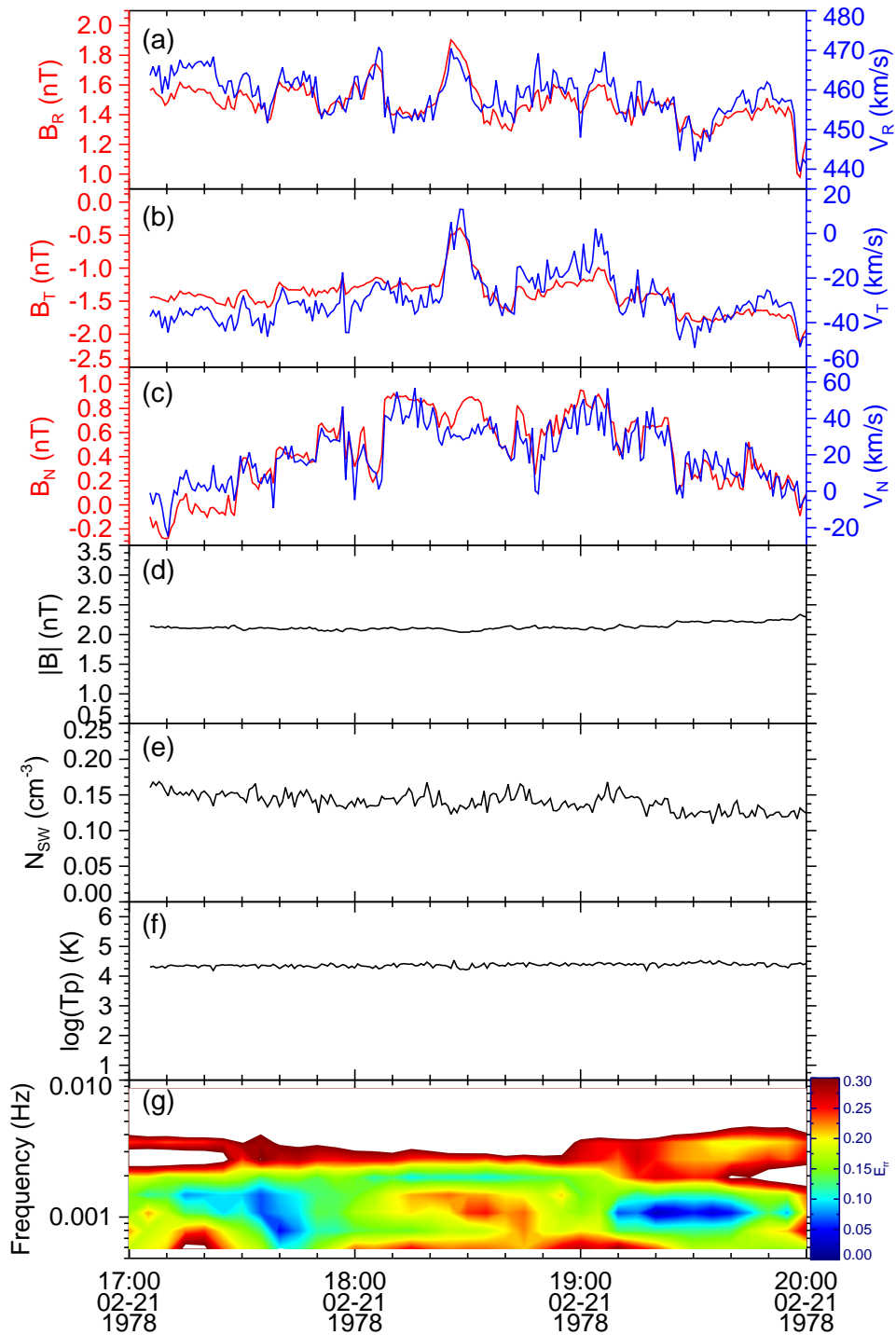}
\caption{Overview of solar wind data on February  21, 1978 observed by \textit{Voyager} 2. (a) Temporal variations of $B_R$ (red) and $V_R$ (blue) in the RTN coordinate system. $B_R$ is generally great than zero, indicating a anti-sunward magnetic sector during that time. (b) Temporal variations of $B_T$ (red) and $V_T$ (blue). (c) Temporal variations of $B_N$ (in red) and $V_N$ (in blue). Panels (d)-(f) gives the magnetic field intensity, the solar wind number density, and the temperature. Panel (e) gives the time-frequency distribution of $E_{rr}$. }
\label{SAF1}
\end{figure}

Figure \ref{SAF1} gives a brief overview of the solar wind properties of a typical SAF event observed by \textit{Voyager} 2 at 2.38 AU between 17:00 UT and 20:00 UT on February 21, 1978. From top to bottom, panels (a)-(f) show the magnetic field ($B_R, B_T, B_N$, in red) and bulk velocity components ($V_R, V_T, V_N$, in blue) in the RTN coordinates, the magnetic field strength ($|B|$), the solar wind proton number density ($N_{SW}$), and solar wind thermal proton temperature ($T_P$), respectively. Panel (g) gives the time-frequency distribution of $E_{rr}$. The background magnetic field during the time interval of this event is estimated to be (1.48, -1.40, 0.44) nT. The background magnetic fields of the upstream and downstream solar wind are (1.45, -1.46, -0.20) nT and (1.37, -1.77, 0.33) nT over the time periods from 14:00--17:00 UT and from 20:00--23:00 UT. The angles between our determinations of the background magnetic fields of the potential SAFs and the upstream and downstream solar wind are 18$^\circ$ and 9$^\circ$, respectively, indicating there is no significant bending of the magnetic field. $B_R$ is great than zero, indicating an anti-sunward magnetic sector during that time. During this time interval, the relative fluctuations of solar wind number density and magnetic field strength are insignificant, of 9\% and 3\%, respectively. However, the three components of $\bf{B}$ and $\bf{V}$ have large-amplitude fluctuations which have a strong positive correlation. The correlation coefficients for the R, T, and N components are 0.62, 0.72, and 0.90, respectively. Such a strong correlation and incompressibility indicate the presence of AFs propagating anti-parallel to the background magnetic field, which strongly suggests a SAF event. The time-frequency distribution of $E_{rr}$ reveals two intervals of relatively pure AFs, from 17:10--18:10 and from 19:00--20:00 UT. The wave period is 810 s -- 1480 s.

\begin{figure}
\centering
\noindent\includegraphics[width=31pc]{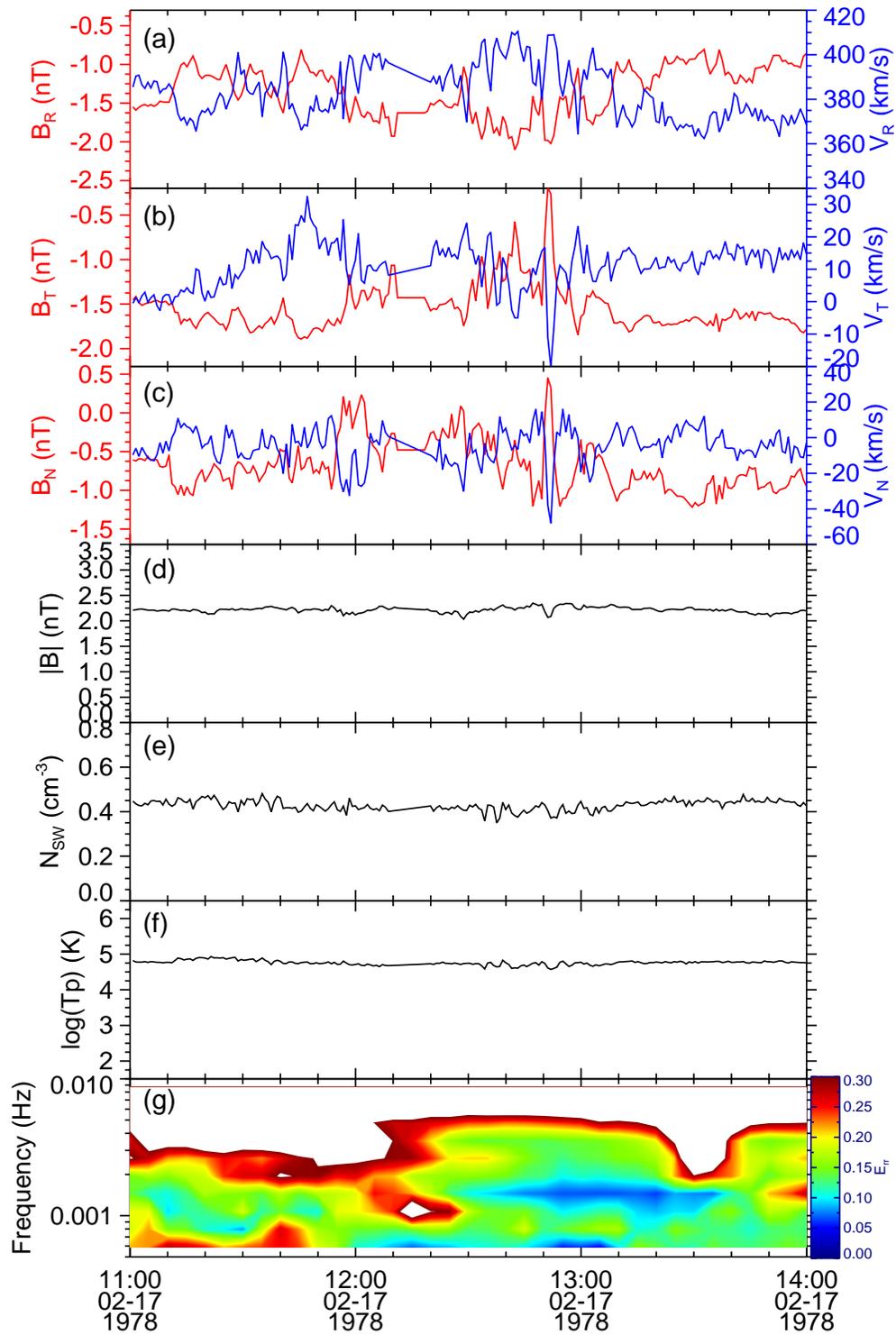}
\caption{Overview of solar wind data on January 19, 1978 observed by \textit{Voyager} 2. The arrangement of the plot is the same as that described in Figure \ref{SAF1}. }
\label{SAF2}
\end{figure}

The AFs observed in the solar wind are not necessarily periodic like a monochromatic wave. Here, they are assumed to be broadband and to propagate in the same direction for each filter. The wave propagation direction is nearly along the background magnetic field, but this direction is hard to determine accurately from a single satellite. The minimum-variance direction obtained from the Minimum Variance Analysis (MVA, \citet{Sonnerup and Cahill 1967}) is assumed to be the wave propagation direction; these two values are used as a double check on the results. If the angle between the wave propagation direction determined by MVA and the background magnetic field was small, the errors in estimating the background magnetic field and determining the wave propagation direction are assumed to be acceptable. For the relatively pure SAFs with periods between 810 s and 1480 s, the correlation coefficients of the fluctuations of magnetic field and solar wind velocity for the R, T, the N components are 0.89, 0.90, and 0.90, respectively. The normal vector of wave propagation direction is calculated to be (-0.811, 0.475, -0.340). As emphasized by \citet{Wang et al 2012}, the ratio of the intermediate to the minimum eigenvalue is an important indicator of the MVA accuracy. For our analysis, that value is as high as 25.0, which confirms the credibility of our MVA results. The angle between the phase velocity direction of the SAFs and the background magnetic field direction ($\theta_{Bn}$) is 165.2$^\circ$. The wave power in the mean field aligned coordinate (MFA) is also calculated. The perpendicular wave power is about 12.7 times larger than the parallel wave power, which indicates the waves are mainly transverse.

Figure \ref{SAF2} gives another SAF event observed by \textit{Voyager} 2 at 2.34 AU from 11:00 to 14:00 UT on February 17, 1978. The background magnetic field is estimated to be (--1.34, --1.53, --0.69) nT. The background magnetic field of the upstream and downstream solar wind are (--0.70, --1.63, --0.84) nT and (--1.13, --1.42, --0.39) nT over the time period from 08:00--11:00 UT and from 14:00--17:00 UT. The angles between the background magnetic fields of the potential SAFs and of the upstream and downstream solar wind are 18$^\circ$ and 7$^\circ$, respectively, indicating there is no significant bending of the magnetic field. $B_R$ is always less than zero, indicating the magnetic sector is sunward. During this time interval, the relative fluctuations of solar wind number density and magnetic field strength are insignificant, of 5\% and 2\%, respectively. However, large-amplitude fluctuations are very clear for all the three components of $\bf{B}$ and $\bf{V}$, with a strong negative correlation between them. The correlation coefficients for the R, T, the N components are --0.94, --0.60, and -0.76, respectively. Such a strong negative correlation and the incompressibility indicate the presence of AFs propagating parallel to the background magnetic field, which indicates a SAF event. The time-frequency distribution of $E_{rr}$ shows relatively pure AFs from 12:20--13:40 UT. The wave period is 600 -- 1450 s. For the relatively pure SAFs with periods between 600 s and 1480 s, the correlation coefficients of the fluctuations for the R, T, the N components are --0.95, --0.84, and --0.96, respectively. The normal vector of the wave propagation direction is (--0.763, --0.644, 0.062). The ratio of the intermediate to the minimum eigenvalue is greater than 4.0, which confirms the credibility of our MVA results. The angle between the phase velocity direction of the SAFs and the background magnetic field direction ($\theta_{Bn}$) is 23.9$^\circ$. The perpendicular wave power is about 7.8 times larger than the parallel wave power, which indicates the waves are mainly transverse.

\begin{figure}[ht!]
\centering
\noindent\includegraphics[width=30pc]{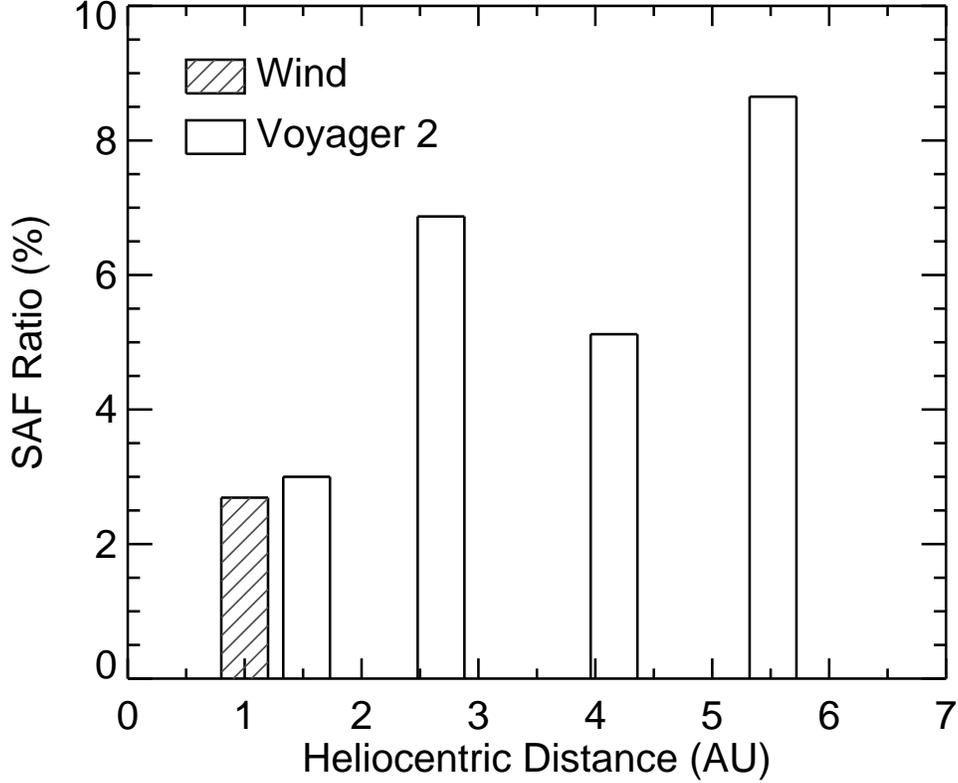}
\caption{Relationship between the ratio of SAFs to AFs versus heliocentric distance. }
\label{ratio}
\end{figure}

\subsection{Dependence of SAF Ratio on Heliocentric Distance}
Figure \ref{ratio} shows the dependence of the SAF ratio on the heliocentric distance. Note that SAFs are a subset of AFs in our analysis. For \textit{Wind} data, the total time durations of SAFs and AFs are 146 hours and 5408 hours, giving the ratio of 2.7\%. In order to make the SAF ratio comparison more reliable in terms of statistical significance, we divide the \textit{Voyager} 2 observations into four time intervals and make sure that the total time durations of AFs in each time interval are almost the same. The heliocentric distance is chosen to be the average value for each interval. For \textit{Voyager} 2 data, the total time durations of AFs for the four intervals are 405.8 hours, 405.7 hours, 405.5 hours, and 407.2 hours. The total time durations of SAFs are 12.2 hours, 31.9 hours, 20.8 hours, and 35.2 hours, respectively. Thus, the ratios are 3.0\%, 6.9\%, 5.1\%, and 8.7\%. The ratio of SAFs to AFs observed by \textit{Voyager} 2 near 1 AU (3.0\%) is very similar to that observed by \textit{Wind} at 1 AU (2.7\%), supporting the validation of our approach. Moreover, the ratio of SAFs to AFs seems to increase with heliocentric distance. If the \textit{Wind} data are divided into four intervals using the same method, the ratios of SAFs to AFs are 2.9\%, 2.4\%, 2.7\%, and 2.8\%, respectively. This indicates that the differences in \textit{Voyager} 2 data are caused by the heliocentric distance changing but not by the data grouping.

The generation mechanism of SAFs is still an open question. As summarized  by \citet{He et al 2015a}, some processes might contribute to the origin of SAFs: (1) AFs are partially reflected in inhomogeneous media, e.g., transverse shear or longitudinal gradients in flow velocity and Alfv\'en speed, and (2) excitation by unstable upstream energetic proton events. The exact generation mechanism of extended trains of SAFs is worthy of future investigation. 

\section{Summary}
\label{d&s}
Sunward-propagating Alfv\'enic fluctuations are believed to be important to heliospheric dynamic processes. However, they have rarely been observed at 1 AU and beyond in the past. We surveyed two years of \textit{Wind} and \textit{Voyager} 2 data before the solar maximum and used the approach proposed by \citet{Li et al 2016} to identify interplanetary Alfv\'enic fluctuations. For \textit{Wind} data at 1 AU, the total time durations of AFs and SAFs are 5408 hours and 146 hours. And for \textit{Voyager} 2 data from 1 AU to 6 AU, the total time durations of AFs and SAFs are 1624 hours and 100 hours, respectively. The occurrence of AFs decreases with heliocentric distance, however, the ratio of SAFs to AFs increases gradually inside 6 AU,  from about 2.7\% at 1.0 AU to about 8.7\% at 5.5 AU. The generation mechanism of extended trains of SAFs is not very clear. New data with high temporal resolution and strahl electron information from future missions will be helpful for understanding this issue more comprehensively.



\acknowledgments

We thank the NSSDC (\url{ftp://nssdcftp.gsfc.nasa.gov}) for access to the data from \emph{Wind} Mission. The high-resolution \emph{Voyager} 2 plasma data used in this study are publicly available at the \emph{Voyager} Data Page of the Massachusetts Institute of Technology Space Plasma Group (\url{ftp://space.mit.edu/pub/plasma/vgr/v2}). The 48 s resolution magnetic field data are accessible at the NSSDC. The authors would like to thank Dr. Shuo Yao for frequent discussions. This work was supported by 973 program 2012CB825602, NNSFC grants 41574169 and 41231067. H. Li was also supported by Youth Innovation Promotion Association of the Chinese Academy of Sciences and in part by the Specialized Research Fund for State Key Laboratories of China.


\begin{thebibliography}{}

\bibitem[\textit{Alfv\'en}(1942)]{Alfven 1942}
Alfv\'en, H. (1942), Existence of electromagnetic - hydrodynamic waves, \textit{\nat, 150}, (3805): 405--406, doi:10.1038/150405d0.

\bibitem[\textit{Bavassano et al.}(2000)]{Bavassano et al 2000}
Bavassano, B., E. Pietropaolo, and R. Bruno (2000), On the evolution of outward and inward Alfv\'enic fluctuations in the polar wind, \textit{\jgr, 105}(A7), 15959--15964, doi:10.1029/1999JA000276.

\bibitem[\textit{Bavassano et al.}(2001)]{Bavassano et al 2001}
Bavassano, B., E. Pietropaolo, and R. Bruno (2001), Radial evolution of outward and inward Alfv\'enic fluctuations in the solar wind: A comparison between equatorial and polar observations by Ulysses, \emph{\jgr, 106}(A6), 10659--10668.
/jgr
\bibitem[\textit{Behannon et al.}(1977)]{Behannon et al 1977}
Behannon, K. W., Acuna, M. H., Burlaga, L. F., Lepping, R. P., Ness, N. F., and Neubauer, F. M. (1977), Magnetic field experiment for Voyagers 1 and 2, \emph{\ssr, 21}(3), 235--257

\bibitem[\textit{Belcher and Davis}(1971)]{Belcher and Davis 1971}
Belcher, J. W., and L. Davis (1971), Large-Amplitude Alfven Waves in Interplanetary Medium, 2., \emph{\jgr, 76}(16), 3534, doi:10.1029/Ja076i016p03534.

\bibitem[\textit{Bridge et al.}(1977)]{Bridge et al 1977}
Bridge, H. S., J. W. Belcher, R. Butler, A. J. Lazarus, A. Mavretic, and J. D. Sullivan (1977), The plasma experiment on the 1977 Voyager mission, \emph{\ssr, 21}, 259--287.

\bibitem[\textit{Bruno et al.}(1997)]{Bruno et al 1997}
Bruno, R., B. Bavassano, E. Pietropaolo, V. Carbone, and H. Rosenbauer (1997), On the inward component of the Alfvenic turbulence in the solar wind, \emph{\jgr, 102}(A7), 14687--14699.

\bibitem[\textit{Burlaga}(1971)]{Burlaga 1971}
Burlaga, L. F. (1971), Hydromagnetic waves and discontinuities in the solar wind, \emph{\ssr, 12}(5), 600--657.

\bibitem[\textit{Burlaga and Turner}(1976)]{Burlaga and Turner 1976}
Burlaga, L. F., and J. M., Turner (1976), Microscale ``Alfv\'en waves" in the solar wind at 1 AU, \emph{\jgr, 81}(1), 73--77.

%

\bibitem[\textit{Gosling et al.}(2009)]{Gosling et al 2009}
Gosling, J. T., D. J. McComas, D. A. Roberts, and R. M. Skoug (2009), A one-sided aspect of Alfvenic fluctuations in the solar wind, \emph{\apjl, 695}(2), L213--L216, doi:10.1088/0004-637X/695/2/L213.

\bibitem[\textit{Gosling et al.}(2011)]{Gosling et al 2011}
Gosling, J. T., H. Tian, and T. D. Phan (2011), Pulsed Alfven waves in the solar wind, \emph{\apjl, 737}(2), L35, doi:10.1088/2041-8205/737/2/L35.


\bibitem[\textit{He et al.}(2011)]{He et al 2011}
He, J., E. Marsch, C. Tu, S. Yao, and H. Tian (2011), Possible evidence of Alfv\'en-cyclotron waves in the angle distribution of magnetic helicity of solar wind turbulence, \emph{\apj, 731}, 85, doi:10.1088/0004-637X/731/2/85.

\bibitem[\textit{He et al.}(2013)]{He et al 2013}
He, J., C. Tu, E. Marsch, S. Bourouaine, and Z. Pei (2013), Radial Evolution of the Wavevector Anisotropy of Solar Wind Turbulence between 0.3 and 1 AU, \emph{\apj, 773}(1), 72, doi:10.1088/0004-637X/773/1/72.

\bibitem[\textit{He et al.}(2015a)]{He et al 2015a}
He, J., Z. Pei, L. Wang, C. Tu, E. Marsch, L. Zhang, and C. Salem (2015), sunward propagating alfven waves in association with sunward drifting proton beams in the solar wind, \emph{\apj, 805}(2), 176, doi:10.1088/0004-637X/805/2/176.

\bibitem[\textit{He et al.}(2015b)]{He et al 2015b}
He, J., C. Tu, E. Marsch, C. H. K. Chen, L. Wang, Z. Pei, Lei Zhang, C. S. Salem, and S. D. Bale (2015), Proton Heating in Solar Wind Compressible Turbulence with Collisions between Counter-propagating Waves, \emph{\apjl, 805}(2), L30, doi:10.1088/2041-8205/813/2/L30.

\bibitem[\textit{Li et al.}(2016)]{Li et al 2016}
Li, H., C. Wang, J. K. Chao, abd W. C. Hsieh (2016), A new approach to identify interplanetary Alfv\'en waves and to obtain their frequency properties, \emph{\jgr, 121}, 42--55, doi:10.1002/2015JA021749.

\bibitem[\textit{Lepping et al.}(1995)]{Lepping et al 1995}
Lepping, R. P., et al. (1995), The wind magnetic field investigation, \emph{\ssr, 71}, 207--229, doi:10.1007/BF00751330.

\bibitem[\textit{Lin et al.}(1995)]{Lin et al 1995}
Lin, R. P., et al. (1995), A three-dimensional plasma and energetic particle investigation for the wind spacecraft, \emph{\ssr, 71}, 125--153, doi:10.1007/BF00751328.

\bibitem[\textit{Marsch}(1991)]{Marsch 1991}
Marsch, E. (1991), MHD turbulence in the solar wind, \emph{In Physics of the inner heliosphere II} (pp. 159-241), Springer Berlin Heidelberg.


\bibitem[\textit{Roberts et al.}(1987)]{Roberts et al 1987}
Roberts, D. A., L. W. Klein, M. L. Goldstein, and W. H. Matthaeus (1987), The nature and evolution of magnetohydrodynamic fluctuations in the solar wind: Voyager observations, \emph{\jgr, 92}(A10), 11021--11040.

\bibitem[\textit{Sonnerup and Cahill}(1967)]{Sonnerup and Cahill 1967}
Sonnerup, B. U. \"{O}, and L. J. Cahill Jr. (1967), Magnetopause structure and attitude from Explorer 12 observations, \emph{\jgr, 72}(1), 171, doi:10.1029/JZ072i001p00171.


\bibitem[\textit{Tu and Marsch}(1995)]{Tu and Marsch 1995}
Tu, C.-Y., and E. Marsch (1995), MHD structures, waves and turbulence in the solar wind: Observations and theories, \emph{\ssr, 73}(1--2), 1--210.

\bibitem[\textit{van der Holst et al.}(2014)]{van der Holst et al 2014}
van der Holst, B., I. V. Sokolov, X. Meng, M. Jin, W. B. Manchester IV, G. T\'{o}th, and T. I. Gombosi (2014), Alfv\'en wave solar model (AWSoM): coronal heating, \emph{\apj, 782}(2), 81.

\bibitem[\textit{V\"{o}lk}(1975)]{Volk 1975}
V\"{o}lk, H. J. (1975), Microstructure of the solar wind, \emph{\ssr, 17}(2--4), 255--276.

\bibitem[\textit{Wang et al.}(2012)]{Wang et al 2012}
Wang, X., J. S. He, C. Y. Tu, E. Marsch, L. Zhang, and J. K. Chao (2012), Large-Amplitude Alfv\'en Wave in Interplanetary Space: The Wind Spacecraft Observations, \emph{\apj, 746}, doi:10.1088/0004-637X/746/2/147.

\bibitem[\textit{Wang et al.}(2015)]{Wang et al 2015}
Wang, X., C. Tu, L. Wang, J. He, and E. Marsch (2015), The upstream-propagating Alfvenic fluctuations with power law spectra in the upstream region of the Earth¡¯s bow shock, \emph{\grl, 42}(10), 3654--3661, doi:10.1002/2015GL063893.

\bibitem[\textit{Yan et al.}(2016)]{Yan et al 2016}
Yan, L., J. He, L. Zhang, C. Tu, E. Marsch, C. H. K. Chen, X. Wang, Linghua Wang, and R. T. Wicks (2016), Spectral Anisotropy of Els\"asser Variables in Two-dimensional Wave-vector Space as Observed in the Fast Solar Wind Turbulence, \emph{\apjl, 816}(2), L24, doi:10.3847/2041-8205/816/2/L24.

\bibitem[\textit{Yang and Chao}(2013)]{Yang and Chao 2013}
Yang, L., and J. K. Chao (2013), Alfv\'en waves in the solar wind, \emph{Chin. J. Space Sci., 33}(4): 353--373.

\end{thebibliography}
\end{document}